# Scaling Analysis of Nanowire Phase Change Memory

Jie Liu, Bin Yu *Fellow IEEE*, and M. P. Anantram

*Abstract*—This letter analyzes the scaling property of nanowire (NW) phase change memory (PCM) using analytic and numerical methods. The scaling scenarios of the three widely-used NW PCM operation schemes (constant electric field, voltage, and current) are studied and compared. It is shown that if the device size is downscaled by a factor of $1/k$ ($k>1$), the operation energy (current) will be reduced by more than $k^3$ ($k$) times, and the operation speed will be increased by $k^2$ times. It is also shown that more than 90% of operation energy is wasted as thermal flux into substrate and electrodes. We predict that, if the wasted thermal flux is effectively reduced by heat confinement technologies, the energy consumed per RESET operation can be decreased from about 1 pJ to less than 100 fJ. It is shown that reducing NW aspect ratio (AR) helps decreasing PCM energy consumption. It is revealed that cross-cell thermal proximity disturbance is counter-intuitively alleviated by scaling, leading to a desirable scaling scenario.

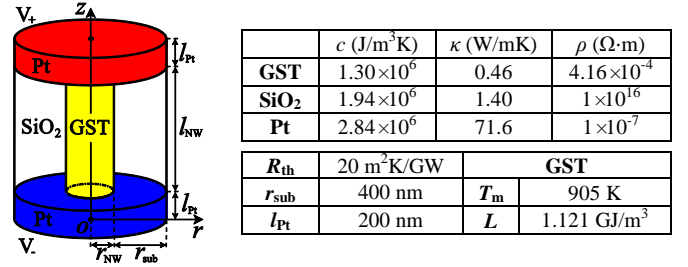

Fig. 1. Schematic NW PCM geometry (left) and default parameters used in the simulation (right) – specific heat capacity $c$, thermal conductivity $\kappa$, electrical resistivity $\rho$, $Ge_2Sb_2Te_5$ (GST) melting point $T_m$, GST latent heat $L$, and thermal boundary resistance $R_{th}$ at GST-Pt interface and GST-SiO$_2$ interface.

## I. INTRODUCTION

THE non-volatile phase change memory (PCM) is promising to replace the flash memory [1]-[2]. As an alternative of the most popular thin film PCM, the novel nanowire (NW) and NW-like pore-shaped PCM researches are gaining momentum in recent years [3]-[9]. However, in these emerging researches, the NW radius, length, aspect ratio (AR), and operation scheme, which all exert significant impacts on device performance, are chosen largely at will and exhibit very large variance. To better understand and harness these influencing factors, and thereby to take advantage of the superior scalability of PCM technology, we present a unified scaling analysis of NW PCM performance by using both analytic and numerical methods. Our analysis reveals the promising scaling scenario by presenting the scaling laws of the crucial physical quantities which determine the PCM device performance. This work is inspired by the temperature scaling analysis in reference [10]. Our focus here is the diffusive transport region, in which the device dimension (tens of nm) is much larger than carrier mean free path $l_{MFP}$ ($l_{MFP}<1$ nm [11]).

Manuscript received June 7, 2011. This work is supported by the U.S. National Science Foundation under Grant Award Number 1006182.
Jie Liu and M. P. Anantram are with the Dept. of Electrical Engr., Univ. of Washington, Seattle, 98195, WA, USA (e-mail: liujie@uw.edu; anant@uw.edu). Bin Yu is with the College of Nanoscale Sci. & Engr., State Univ. of New York, Albany, NY, USA (email: byu@uamail.albany.edu).

## II. ANALYTIC SCALING ANALYSIS

NW PCM research is still at its early stage and the device operation schemes are diverse. To perform the RESET operations, the existing NW PCM work use either constant current [4]-[6] or constant voltage [7]-[9] pulses. Also, it was shown that the threshold switching of SET operations exhibit constant field scaling [6]-[7]. In this section, we will provide a general analysis and comparison of the scaling behaviors in the constant electric field, voltage and current operation schemes.

In the cylindrical coordinate as shown in Fig. 1, the governing equations of electro-thermal transport in NW PCM before and after isotropic scaling are

$$\rho J' = E' = -\nabla'V' = -\nabla V \alpha = E \alpha = \rho J \alpha \quad (1)$$

$$c\, \partial_t T - \kappa\, r^{-1}\, \partial_r(r\partial_r T) - \kappa\, \partial_{zz}T - \rho J^2$$
$$= c\, \partial_{t'}T\, \beta - \kappa\, r'^{-1}\, \partial_{r'}(r'\partial_{r'}T)/k^2 - \kappa\, \partial_{z'z'}T/k^2 - \rho J'^2/\alpha^2 \quad (2)$$

where $r'=r/k$, $z'=z/k$ ($k>1$), $t'=t\beta$. $\alpha$ is determined by how the electric field behaves with scaling in Eq. (1). $\beta$ is chosen to keep Eq. (2) invariant after scaling. Here, we use the variables without (with) prime to denote quantities before (after) scaling. The meanings of the symbols are listed in Fig. 1 and Tab. I. We observe that: (i) If electric field is kept constant in scaling ($E'=E$), Eq. (1) requires $\alpha=1$ and $V'=V/k$; Eq. (2) is invariant if $T'=T/k^2$ and $\beta=1/k^2$. As typical pulse duration is long enough for temperature to reach its steady state, making the pulse width longer than $t'=t/k^2$ will not increase the temperature beyond $T'=T/k^2$ and the PCM cell can never be switched to the RESET state. (ii) If voltage is kept constant in scaling ($V'=V$), Eq. (1) requires $\alpha=k$; Eq. (2) is invariant if $T'=T$ and $\beta=1/k^2$. This indicates that the temperature amplitude does not change but the time required to reach the same amplitude is reduced by $k^2$ times [10]. (iii) If current is kept constant in scaling ($I'=I$, so $J'=Jk^2$),

Eq. (1) requires $\alpha=k^2$ and $V'=Vk$; Eq. (2) is invariant if $T'=Tk^2$ and $\beta=1/k^2$. This means that the temperature amplitude becomes $k^2$ times larger than that before scaling even the pulse duration is reduced by $k^2$ times. Special attention should be paid to the sharply increase electric field $E'=Ek^2$, because this intensifies electro-migration, which limits the lifetime of device.

The scaling factors are summarized in Tab. I. We can see that, in all of the three cases, $t$ is scaled by a factor of $1/k^2$, indicating that the changing speed of $T$ is $k^2$ times faster in the scaled PCM. So, the RESET speed will be increased by $k^2$ times. The SET speed, however, is largely determined by crystallization speed, instead of changing speed of $T$. Therefore, although experiment has shown that SET time can be reduced from about 100 ns to less than 10 ns if PCM cell size is scaled from 500 nm to less than 50 nm [3], a microscopic theoretical analysis is required to quantitatively define the impact of scaling on SET speed. Since RESET and SET operations adopt the constant voltage [7] and constant field [12] scaling, the RESET and SET energy $Q$ (current $I$) scaling factors are $1/k^3$ ($1/k$) and $1/k^5$ ($1/k^2$), as shown in Tab. I.

TABLE I. ANALYTIC SCALING FACTORS

| | Quantity | Const. $E$ | Const. $V$ | Const. $I$ |
|---|---|---|---|---|
| $r$ | radius | $1/k$ | $1/k$ | $1/k$ |
| $l$ | length | $1/k$ | $1/k$ | $1/k$ |
| $R$ | resistance | $k$ | $k$ | $k$ |
| $E$ | electric field | 1 | $k$ | $k^2$ |
| $V$ | voltage | $1/k$ | 1 | $k$ |
| $I$ | current | $1/k^2$ | $1/k$ | 1 |
| $J$ | current density | 1 | $k$ | $k^2$ |
| $T$ | temperature | $1/k^2$ | 1 | $k^2$ |
| $t$ | time | $1/k^2$ | $1/k^2$ | $1/k^2$ |
| $P$ | power | $1/k^3$ | $1/k$ | $k$ |
| $Q$ | energy | $1/k^5$ | $1/k^3$ | $1/k$ |

III. NUMERICAL SCALING ANALYSIS

The aforementioned analytic scaling analysis reveals the benefits of scaling to reduce energy consumption and to increase device operation speed. However, firstly it did not include the latent heat of melting; and secondly it scaled the thermal boundary resistance (TBR) $R_{th}=-\Delta T/\kappa \nabla T$ artificially to $R'_{th}=R_{th}/k$. Here, $\Delta T$ means the temperature discontinuity ($\Delta T>0$) at material interface. To overcome these drawbacks and obtain a more accurate scaling analysis, we numerically solve the Laplace equation $\nabla^2 V=0$ to obtain current density $J=-\rho^{-1}\nabla V$. Then heat equation (Eq. (2)) is solved for temperature distribution by using the time dependent finite element method (TD-FEM). In solving Laplace's equation, the Dirichlet boundary condition $V_+$ and $V_-$ are applied at top and bottom Pt surface, as shown in Fig. 1. In solving the heat equation, the Dirichlet boundary condition $T=300$ K is imposed at top and bottom Pt surface. At $r=0$ and at $r=r_{NW}+r_{sub}$, the homogeneous Neumann boundary conditions are applied for both $T$ and $V$. The $V_+$ and $V_-$ are chosen so that the maximum value of $T(r_{NW},z,t_R)$ exceeds $T_m$, because this ensures one cross-section of NW is melted within pulse duration time $t_R$ and then quenched to amorphous state at $t>t_R$ when the pulse is off.

By using the method outlined above and parameters shown in Fig. 1, the scaling behaviors of crucial physical quantities that determine NW PCM RESET performance are shown in Fig. 2-3. In these plots, $r'_{NW}=40/k$ nm; $l'_{NW}=r'_{NW}\times AR$ nm; and $\Delta t'=20/k^2$ ns. The four components (radial heat loss into $SiO_2$ substrate $Q_{rad}$, axial heat loss into Pt electrodes $Q_{axi}$, energy used due to heat capacity $Q_{cap}$, and energy used due to latent heat $Q_{lat}$) that determine RESET energy $Q$ are plotted in Fig. 2-3 (c). Here, $Q_{rad}$ ($Q_{axi}$) is obtained by integrating the thermal flux at the GST-$SiO_2$ (GST-Pt) interface. It is obvious that more than 90% of the RESET energy is wasted as thermal energy loss into substrate and electrodes. For example, for NW with $r_{NW}=l_{NW}=20$ nm (leftmost points of Fig. 2-3 with AR=1), $Q\approx 0.64$ pJ, which consists of $Q_{rad}\approx 0.27$ pJ, $Q_{axi}\approx 0.32$ pJ, $Q_{cap}\approx 0.04$ pJ, and $Q_{lat}\approx 0.01$ pJ. Actually, only $Q_{cap}+Q_{lat}\approx 50$ fJ is necessary to achieve the RESET operations. The wasted $Q_{rad}$ and $Q_{axi}$ consume 92% of $Q$. So, confining heat during RESET operation has the potential to improve energy efficiency by one order of magnitude. To do this, one can either use substrate and electrodes with small effective $\kappa$ or use materials which have large TBR with chalcogenides (see inserts of Fig. 2-3(c)). Our simulation reveals that if $\kappa$ values of the Pt and $SiO_2$ regions are decreased by 2 orders of magnitude, the RESET energy (current) can be reduced by 60%-70% (40%-50%).

To more clearly reveal the quantitative scaling relations, the numerical scaling factors are listed in Tab. II. From Fig. 2-3(b) and Tab. II, we can see that RESET operation adopts constant voltage scaling roughly. The numerical scaling factors of $E$, $V$, $I$, $J$, $P$, and $Q$ are slightly smaller than the corresponding analytic scaling factors shown in Tab. I. So, Tab. I is a lower limit of the improvement in device performance during scaling. We have shown that in constant voltage scheme, $T'=T$. This means that the thermal proximity effect is kept the same during scaling. However, this conclusion is based on $R'_{th}=R_{th}/k$. In the spatial scale of our interest (tens of nm), we expect $R'_{th}\approx R_{th}>R_{th}/k$. So, downscaling will alleviate the thermal proximity disturbance.

Fig. 2-3 and Tab. II tell us that the above scaling conclusions are valid for various AR values. Fig. 4 reveals that decreasing the AR reduces the energy required but at the cost of increasing $I$. As large $I$ limits the cell selector size and, hence, the data density, we need to strike a balance between energy efficiency and data density to select an appropriate AR value.

In this work, $\kappa$ and $T_m$ are chosen to be the bulk values. In the nm-scale, $\kappa$ ($T_m$) can be suppressed due to modified phonon dispersion (interfacial energy [4][9]). So, the scaling scenario will be even better than those listed in Tab. II and Fig. 2-3.

In our analysis, we have neglected barriers (Schottky or other barriers) at Pt-NW interface, which can modify heat dissipation leading to hot spots. The temperature dependence of resistivity is also neglected. These issues deserve future investigation.

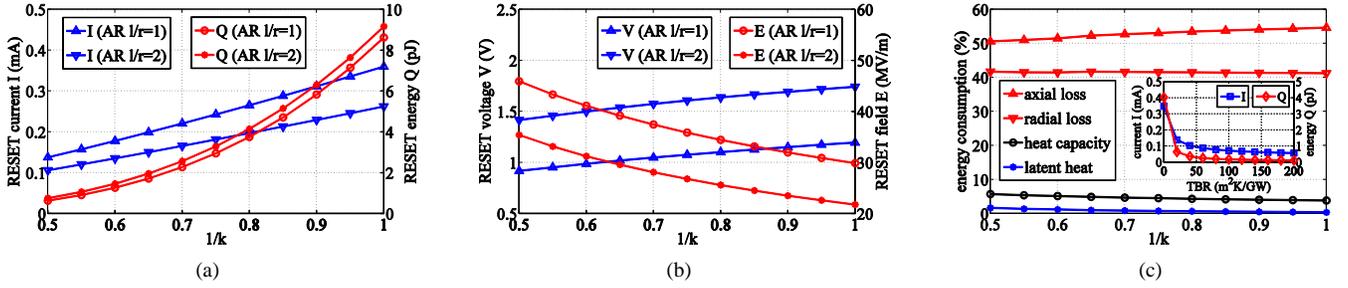

Fig. 2. RESET performance scaling of NW PCM with small AR ($l_{NW}/r_{NW}$) using TD-FEM. In (c), AR=1.

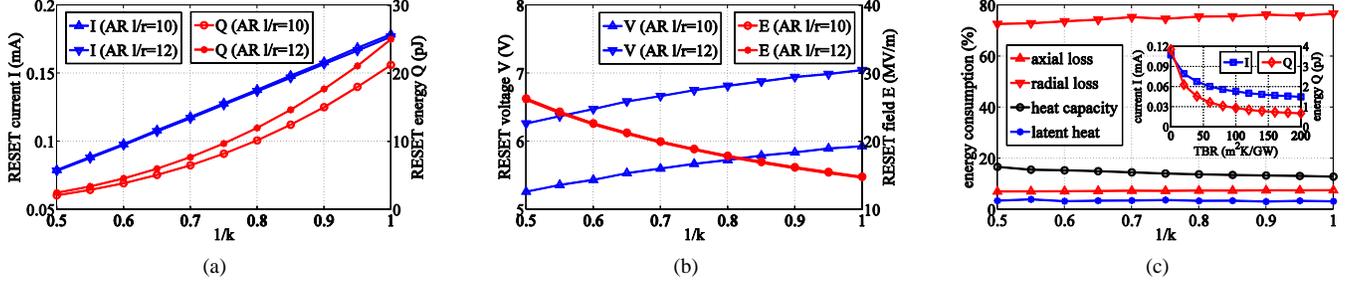

Fig. 3. RESET performance scaling of NW PCM with large AR ($l_{NW}/r_{NW}$) using TD-FEM. In (c), AR=10.

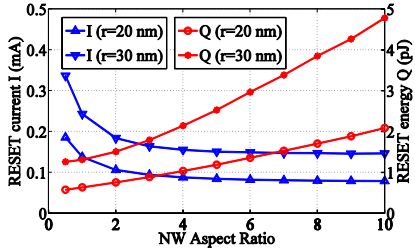

Fig. 4. Impact of AR ($l_{NW}/r_{NW}$) on RESET performance.

TABLE II. NUMERICAL SCALING FACTORS ($k=2$)

| | Numerical Scaling Results | | | | | | Analytic Scaling (Const. V in Tab. I) |
|---|---|---|---|---|---|---|---|
| | Small AR (Fig 2) | | Medium AR | | Large AR (Fig. 3) | | |
| | AR 1 | AR 2 | AR 5 | AR 6 | AR 10 | AR 12 | |
| $r$ | 0.500 | 0.500 | 0.500 | 0.500 | 0.500 | 0.500 | 0.500 |
| $l$ | 0.500 | 0.500 | 0.500 | 0.500 | 0.500 | 0.500 | 0.500 |
| $R$ | 2.000 | 2.000 | 2.000 | 2.000 | 2.000 | 2.000 | 2.000 |
| $E$ | 1.537 | 1.626 | 1.721 | 1.740 | 1.775 | 1.778 | 2.000 |
| $V$ | 0.769 | 0.813 | 0.861 | 0.870 | 0.887 | 0.889 | 1.000 |
| $I$ | 0.383 | 0.406 | 0.429 | 0.434 | 0.443 | 0.443 | 0.500 |
| $J$ | 1.539 | 1.627 | 1.722 | 1.741 | 1.774 | 1.778 | 2.000 |
| $T$ | 1.000 | 1.000 | 1.000 | 1.000 | 1.000 | 1.000 | 1.000 |
| $t$ | 0.250 | 0.250 | 0.250 | 0.250 | 0.250 | 0.250 | 0.250 |
| $P$ | 0.296 | 0.331 | 0.370 | 0.379 | 0.393 | 0.395 | 0.500 |
| $Q$ | 0.074 | 0.083 | 0.093 | 0.095 | 0.098 | 0.099 | 0.125 |

## I. CONCLUSION

A unified scaling analysis of the NW PCM under constant electric field, voltage, and current conditions is presented. The impact of isotropic downscaling of device size by a factor of $1/k$ ($k>1$) is investigated. We show that for RESET operation (i) constant electric field scaling is not a viable region of operation, (ii) constant voltage scheme is desirable, (iii) scaling alleviates cross-cell thermal proximity disturbance, (iv) operation speed is increased by $k^2$ times, and (v) operation energy (current) is reduced by more than $k^3$ ($k$) times. It is demonstrated that more than 90% of the operation energy is wasted as thermal flux into substrate and electrodes. If heat can be effectively confined, the energy consumed per RESET operation can be reduced from about 1 pJ to less than 100 fJ.